# Polarization-Dependent Resonant Anomalous Surface X-ray Scattering of CO/Pt(111)

Andreas Menzel,[1] Yuriy V. Tolmachev,[1a] Kee-Chul Chang,[1] Vladimir Komanicky,[1] Yong S. Chu,[2] John J. Rehr,[3] and Hoydoo You[1]

[1] Argonne National Laboratory, Materials Science Division, 9700 S. Cass Ave., Argonne, IL 60439

[2] Argonne National Laboratory, Experimental Facility Division, 9700 S. Cass Ave., Argonne, IL 60439

[3] University of Washington, Department of Physics, P.O. Box 351560, Seattle, WA 98195

Polarization dependence of resonant anomalous surface x-ray scattering (RASXS) was studied for interfaces buried in electrolytes or in high-pressure gas. We demonstrate that RASXS exhibits strong polarization dependence when the surface is only slightly modified by adsorption of light elements such as carbon monoxide on platinum surfaces. $\sigma$- and $\pi$- polarization RASXS data were simulated with the latest version of *ab initio* multiple scattering calculations (FEFF8.2). Elementary considerations are additionally presented for the origin of the polarization dependence in RASXS.

61.10.-i, 68.35.Ct, 82.45.Jn

---

[a] Current address: Kent State University, Chemistry Department, P.O. Box 5190, Kent, OH 44242

Anomalous x-ray scattering is a virtual absorption/emission process of photon probing electronic transitions between core levels of the resonant atom and unoccupied states near the Fermi level. It is a powerful tool for the investigation of electronic and chemical states of the resonant atom. It exhibits generally strong polarization dependence when the resonant atom is in an anisotropic environment. This was originally demonstrated on $K_2PtCl_4$ single crystals with a highly anisotropic layered symmetry by tuning x-ray energy through the Pt $L_{III}$ edge.[1,2] Since then, polarization dependent resonant scattering, also referred to as Templeton scattering, has been applied to many bulk materials, e.g., in the study of anisotropic orbital ordering in oxide materials.[3]

The availability of powerful synchrotron x-rays now allows to apply the resonance technique to surface x-ray scattering (SXS). We refer to this technique as resonant anomalous surface x-ray scattering (RASXS). It has been found particularly useful for detecting monolayer-level structure and chemical changes of buried interfaces such solid-solid interfaces[4,5] and solid-liquid interfaces.[6,7,8] In this letter, we will show that the effect of Templeton scattering, namely, a strong polarization-dependent scattering, can be clearly seen in RASXS. Analogous to diffraction anomalous fine structure (DAFS),[9] we will demonstrate that spectra obtained in RASXS measurements can be compared to and eventually analyzed by the latest *ab initio* multiple-scattering calculations (FEFF8.2)[10] to yield local and chemical information heretofore unavailable from buried interfaces.

Recently, Chu, *et al.* have shown that the chemical environment of platinum surface oxidation can be studied with RASXS.[7] The RASXS measurements were performed with σ-polarized synchrotron x-rays for geometric convenience. More recently, we have found that scans made with π-polarization are qualitatively different from those with σ-polarization.[11] The most notable difference is the sharply rising peak several eV above the edge which is absent in case of the non-oxidized surface. We attributed the peak to the large overlap of the π-polarization with Pt$d$-O2$p$ bonds in the transition matrix elements. This convinced us that the polarization dependence is easily measurable and that it may yield important chemical information of the interface.

In this letter, we report the first successful modeling of polarization-dependent RASXS spectra. We will not focus on Pt surface oxidation because it involves the restructuring of the surface platinum atoms[7] which complicates a clean demonstration of the polarization dependence. Instead, we report on the measurements and modeling performed for (2×2)-3CO close-packed dense CO monolayers on Pt(111) surfaces which we have recently studied both in CO-saturated solution[12] and in ~1 atm CO gas.[13] The system of CO/Pt(111) is generally considered a standard system in surface science and important from the fundamental scientific point of view. We will focus on the Pt $L_{III}$ edge which was proven to be chemically sensitive.[2,7,8]

Experiments were performed at beam line 11-ID-D at the Advanced Photon Source (APS), Argonne National Laboratory. The monochromator of the beamline is cryogenically cooled Si(220) with an energy resolution of ~1.0 eV. The incoming x-rays were linearly polarized in the horizontal plane. A six-circle diffractometer[14] was used to constrain the surface normal either to the vertical plane for σ-polarization or to the horizontal plane for π-polarization

Two types of background subtraction were always employed in order to ensure that only elastically scattered x-rays were detected. First, we used energy-dispersive detectors, such as Si drift diode detectors (Radiant Detector Technologies at Northridge, CA USA). The detector signals were fed to an amplifier typically with 0.25 or 0.125 μsec shaping time and base-line restoration, and the amplifier output subsequently fed to a pulse height analyzer (AIM module by Canberra at Meriden, CT USA). In this way, we were able to discriminate most of the fluorescence background. The shaping time was chosen to compromise the high count rate and suitable energy resolution, ~180 eV, to discriminate the fluorescence background scattering. Second, we remove diffuse background for every data point of the scans by subtracting the duplicated measurements with sufficient angular offset of ~0.5° from the surface peak. When both types of background subtractions are done only the true elastically scattered x-rays from the surface or interface are counted.

A ball model of the (2×2)-3CO on Pt(111) is shown in Figure 1(a). Open circles indicate Pt atoms and solid circles CO molecules. Figure 1(b) shows the reciprocal lattices of the surface. Large red circles mark Pt reflections and small circles strong CO peaks. SXS measurements[12,13] indicate a close-packed triangular CO lattice. Each CO molecule bonds with the C sides down[15] to atop, hcp, and fcc sites of Pt(111) surface.

In resonance condition, we find a pronounced difference depending on whether x-ray polarization is in-plane (σ) or out-of-plane (π) in the presence of CO on the surface. In Figure 2, we show three RASXS scans though the Pt $L_{III}$ edge taken at the anti-Bragg position (½ ½ ½): a π-polarization scan with CO present on the surface (black squares), a σ-polarization scan with CO (blue triangles), and a π-polarization scan without CO (red circles). Without CO on the surface, the σ-polarization scan is essentially similar to the π-polarization scan and, therefore, not shown. The scattering geometries are shown in the inset. The polarization of the incoming x-rays does not change but the surface orientation changes from 'near horizontal' (top) to 'vertical' (bottom) for σ- and π-polarizations, respectively. We have performed similar measurements on the (2×2)-3CO gas phase, which closely reproduce the in-electrolyte data that are presented here. Further, we took polarization-dependent RASXS measurements on CO superstructure peaks, e.g., (½ ½ 0.2) in the surface hexagonal unit defined in Figure 1(b). These measurements are being analyzed and the results will be published elsewhere.

The scans presented here were measured in 0.1 M sulfuric acid *in situ* under electrochemical potential control at -200 mV vs. Ag/AgCl.[12] The solid line is the atomic Cromer-Liberman calculation.[16] The raw data, $I\left(\frac{1}{2}\frac{1}{2}\frac{1}{2};E\right)$, were fit to the calculation below 11.55 keV and above the 11.65 keV in order to estimate the non-resonant intensity, $I_0 = (f_0/2)^2$, where $f_0$ is the Thomson term. The factor 2 comes from the half-monolayer scattering at the anti-Bragg condition.[7] Then the normalized intensity can be expressed as

$$I(\tfrac{1}{2}\tfrac{1}{2}\tfrac{1}{2};E)/I_0 = 4\left|\tfrac{1}{2}(f_0 + f_R) + f_{R-CO} + f_{CO}\right|^2 / f_0^2$$
$$= 1 + 2\operatorname{Re}[f_R]/f_0 + 4\operatorname{Re}[f_{R-CO}]/f_0 + \cdots$$

where $f_R$ is the unmodified Pt resonance term, $f_{R-CO}$ the resonance term induced by CO adsorption, and $f_{CO}$ the non-resonant scattering contribution of the CO layer. All higher order terms and non-resonant terms are not shown for clarity.

Some difference between the two σ-polarization scans can be seen. The resonance dip goes deeper when the CO layer is present. Otherwise, the two spectra are quite similar. However, the π-polarization scan with the CO layer (black squares) is significantly different from these two scans. The sharply rising feature several eV above the resonance energy is unique for the π-polarization scan with CO absorbed on the surface. It is highlighted in Figure 2(b) as difference of the normalized intensity, $NI^{\pi}_{Pt-CO} - NI^{\sigma}_{Pt}$ (squares) and $NI^{\sigma}_{Pt-CO} - NI^{\sigma}_{Pt}$ (triangles). We speculate that the excited Pt core electron experiences a strongly anisotropic environment due to the empty states present in Pt-C and C-O bonds. Since no such feature can be seen in both polarizations without CO, we are confident that it is purely due to the adsorption of CO.

One approach to interpret the scans shown in Figure 2 is first-principle calculations where one computes the density of states and the corresponding transition matrix elements. Such calculations are currently beyond the scope of this letter. However, the following qualitative discussion will help us to understand the origin of the polarization dependence. We have shown that one can separate the CO specific transition matrix elements from all other resonance terms.[11] The sensitivity of the π-polarized x-rays to the surface bonding is the result of overlap integrals similar to the polarization dependence well established for soft x-ray absorption.[17] In our case, however, the excited electrons are not of the adsorbates but the substrate. Electronic configurations of CO are well known and several unoccupied states, including the $2\pi^*$ state, are within ~5 eV above the Fermi level.[18] Since these states are narrow, a bipolar resonance behavior, $1/(E-E_0)$ where $E$ and $E_0$ are the x-ray energy and the resonance energy, respectively, can be expected. We believe that the bipolar behavior seen in the difference counts [Figure 2(b)] near the Pt $L_{III}$ resonance energy are evidence for the extra density of states due to the presence of CO on the surfaces.

Another approach is to compute the resonant scattering factors by calculating multiple-scattering of the excited core electron by nearby atoms before being reabsorbed. Compared to first-principle calculations, the multiple-scattering approach is currently more practical and has been significantly developed over the last decade.[10] For this reason, we will use the multiple-scattering approach for simulations of the scattering spectra and compare to our data. We have employed the FEFF8.2 code, the latest version of the series,[10] in calculating the scattering intensities normalized in the same manner as the data shown Figure 2. Within the current version of the code, the calculations are limited to forward scattering. We do not expect that this limitation will significantly affect the main features of our results. We used clusters containing only nearest or next-nearest neighbors to the excited atom and limited the multiple scattering path to 7.5 Å.

We used the structural parameters of the clusters determined in SXS measurements.[12,13] The results are shown in Figure 3. Calculations with larger clusters and path lengths produce essentially similar results (not shown) with some differences only in details. The clusters used in the computation are shown in the inset. Open circles represent Pt atoms, and blue and red circles carbon and oxygen atoms, respectively. Note that, per unit cell, one CO molecule is linearly bonded to a Pt atom and two others are bonded to three Pt atoms by occupying 3-fold hollow sites. In the computations, $L_{III}$ ($2p_{3/2}$) electrons are excited. In case of the upper cluster, the center Pt atom bonded to atop CO is excited while, in the case of the lower cluster, the center Pt atom bonded to hollow-site CO is excited. Only the π-polarization results are shown in Figure 3. Pt directly under atop CO exhibits the most noticeable feature right above the edge. Pt atoms under hollow-site CO show a similar feature albeit less pronounced. The weighted average of one Pt under atop CO and three Pt atoms under hollow-site CO is also shown. Although there are some differences between the data shown in Figure 2 and the calculations shown in Figure 3, the main features are well reproduced.

A more direct comparison of our data to the computation is displayed in Figure 4 by overlaying the calculations over the difference between π- and σ-polarization scans in the normalized counts, ($NI_\pi$-$NI_\sigma$). In this way, the data and computations can be compared without further adjustable parameters beyond the normalization constants. The red dashed and green dot-dashed lines are the calculations made for atop and 3-fold Pt sublattice atoms, respectively, the blue solid line is the weighted combination of the (2×2)-3CO structure. The agreement is acceptable since the solid curve tracks the data quite well despite some discrepancies in the amplitudes. We anticipate that the agreement should improve as our computation scheme improves to allow fitting the structure parameters.

We can make a crude interpretation of the oscillating features in the spectra as multiple scattering of the electron excited by x-rays. The wavelength of an electron at ~ 6 eV above the resonance energy, where we find the sharply rising feature, is ~ 5 Å. This is approximately twice the distance between surface Pt and center-of-mass of CO.[12,13] Since the average center positions of the unoccupied C-O bonds can be estimated to be the center of mass, one can imagine a standing wave formation of electron between two attractive potential centers at Pt and CO layers. This resonant standing wave formation can explain the sharply rising feature in the spectra. Similarly, we can assign the peaks at 27 eV (2.4 Å), 46 eV (1.8 Å), 71 (1.5 Å), and 83 eV (1.3 Å) to the average distance between Pt and CO, that of Pt and C of atop CO, that of Pt and C of 3-fold CO, and that of C and O, respectively.[12,13]

In conclusion, we have shown strong polarization dependence of the elastic surface-sensitive scattering from buried interfaces. We attribute this polarization dependency to the anisotropic *local* environment of Pt surface atoms to which CO bonds. The polarization-dependent RASXS measurements like this have the potential to be used to study local and chemical information of the buried interfaces with light elements using hard x-rays.

We would like to thank Drs Guy Jennings and Klaus Attenkofer for useful discussion and assistance during the experiments. This work and the use of the Advanced Photon Source were supported by the Office of Basic Energy Sciences, U.S. Department of Energy under contract no. W-31-109-ENG-38.

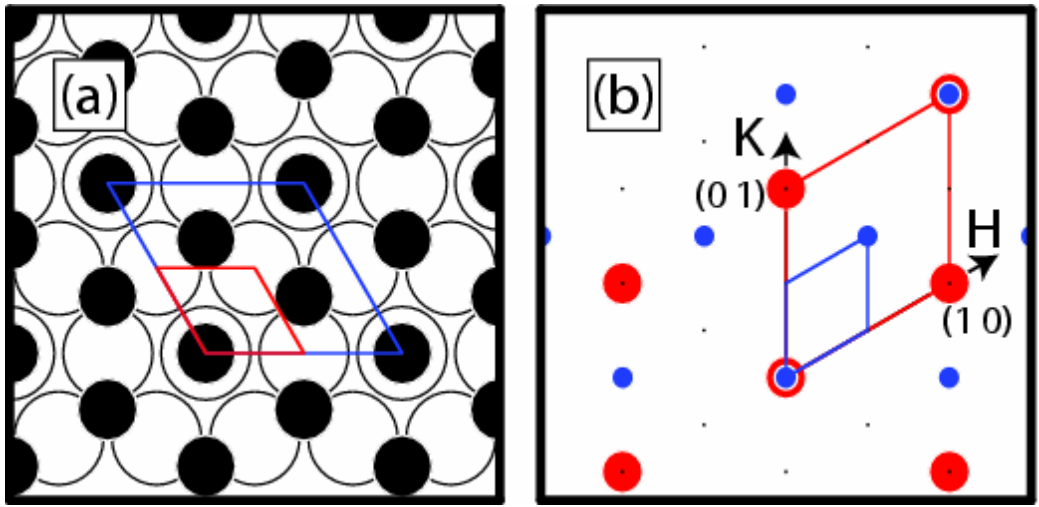

Figure 1. (a) A ball model of (2×2)-3CO on Pt(111). (b) The corresponding reciprocal lattices. The unit cells are shown by parallelograms.

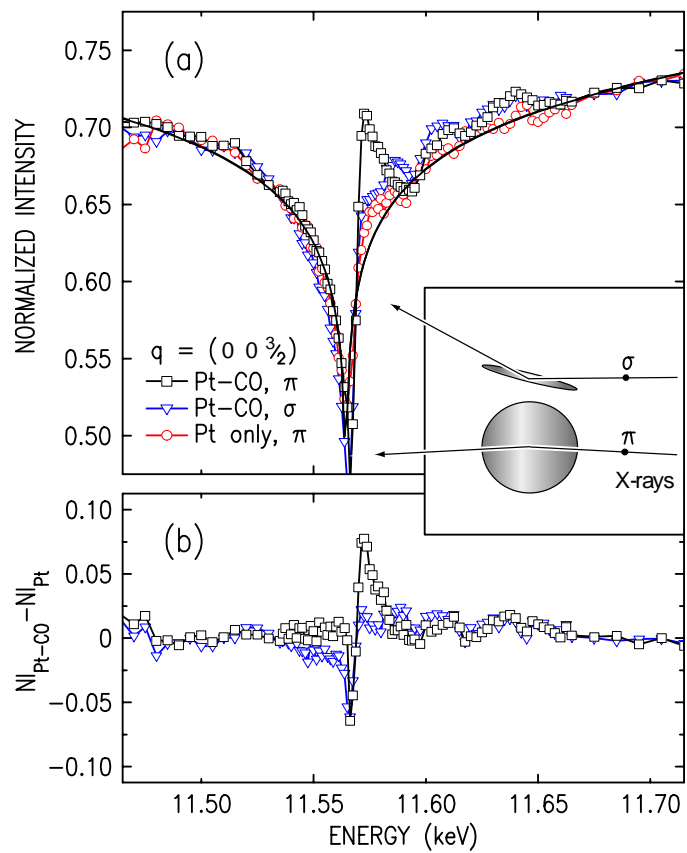

Figure 2. Scans through Pt $L_{III}$ edge. (a) Normalized intensities for π- and σ- polarizations with CO and π-polarization without CO. The solid line is the Cromer-Liberman calculation.[16] (b) Differences in the normalized intensities due to CO absorption.

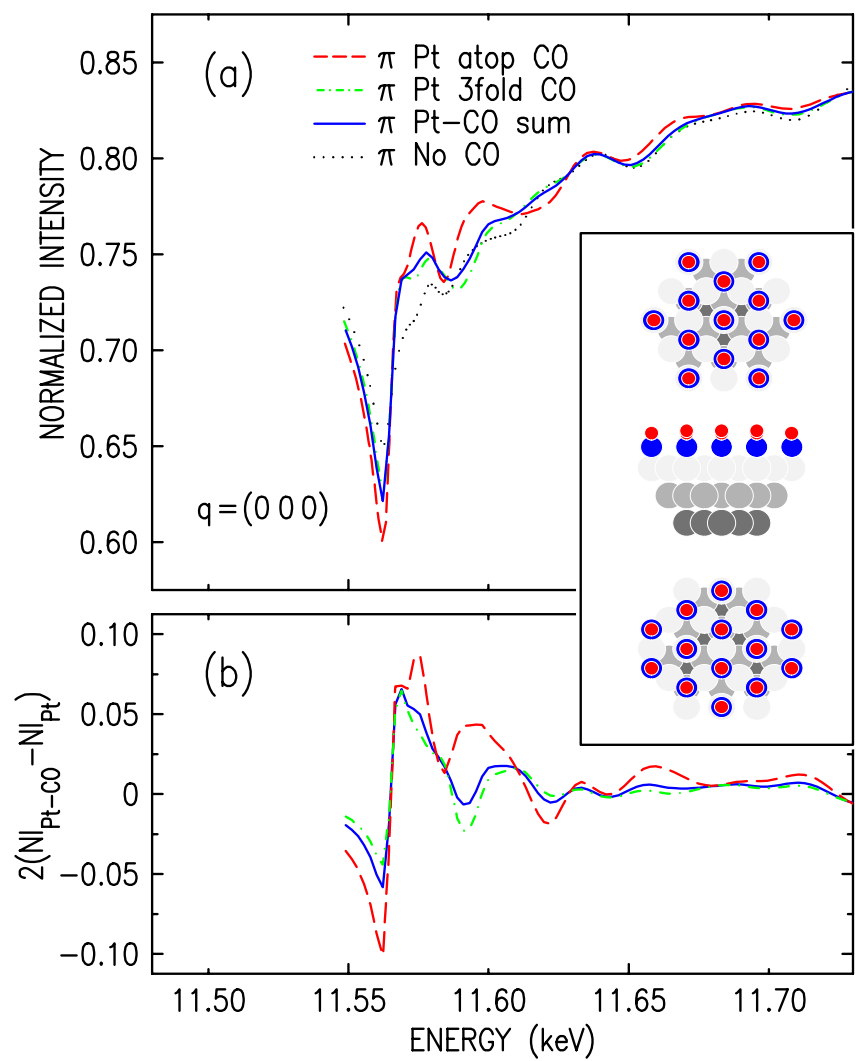

Figure 3. FEFF8.2 calculations for various cases (a) and the difference from the Pt-only case (b). The clusters used in the calculation are shown in the inset. The center Pt atoms in the surface (top views) were excited in the calculations.

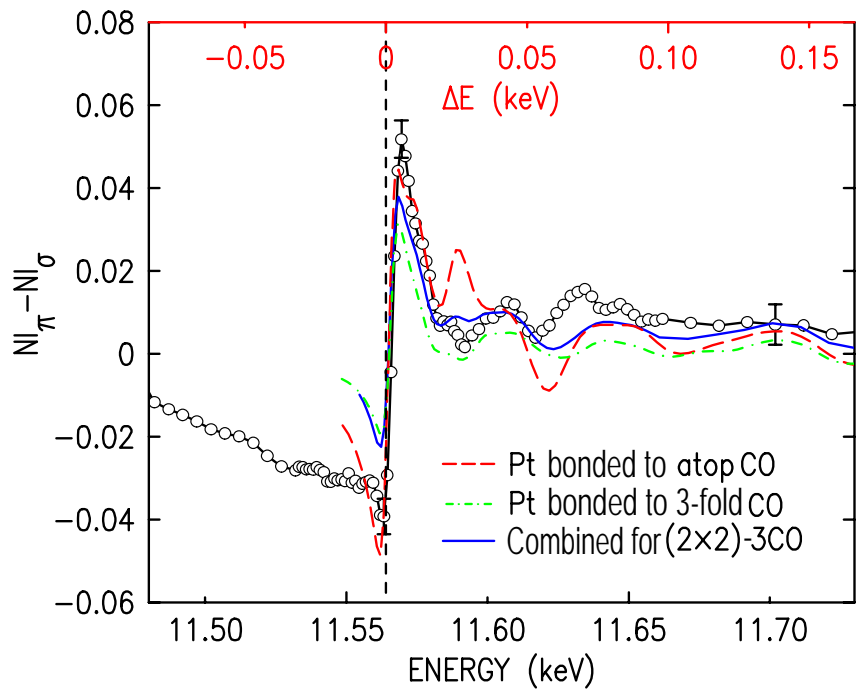

Figure 4. Difference counts of the experimental data are compared to FEFF calculations as described in the text.